\documentclass[12pt]{article} 
\usepackage{amssymb}
\begin{document} 
\begin{center}
          {\large \bf The wake in mid-central nuclear collisions } 

\vspace{0.5cm}                   
{\bf I.M. Dremin}

\vspace{0.5cm}              
          Lebedev Physical Institute, Moscow 119991, Russia

\end{center}

\begin{abstract}
It is argued that PHENIX collaboration observed for the first time the radiation 
of the longitudinal wake oscillations formed behind the parton penetrating
the quark-gluon medium. It shifts the maximum of a hump in two-particle 
correlations and changes its width in the case of some special orientation
of the trigger particle.
\end{abstract}


The quark-gluon medium formed for a very short time interval in the collisions 
of high energy nuclei possesses collective properties. This has been 
observed by its response to high-energy partons passing through it. In 
particular, the spectacular two-hump structure of two-particle correlations
near the direction of away-side jets in central collisions was interpreted
\cite{drem1} as color Cherenkov radiation. The chromopermittivity of the medium
was found \cite{dklv} by comparison of theoretical predictions with 
experimental data. New experimental results \cite{holz} on mid-central nuclear
collisions point out to another collective effect being observed. It has been 
found that the similar two-hump structure appears but with some additional 
contribution noticeable at a particular orientation of the trigger particle
just in between the in-plane and out-of-plane directions. In his presentation 
of the PHENIX collaboration results W.G. Holzmann states that "at present, it is 
unclear whether this merely reflects a geometry dependent shift in the away-side  
 peaks or perhaps an additional contribution at $\Delta \phi =\pi /2$".
Here, I'd like to argue that this is an additional contribution in the vicinity
of $\Delta \phi = \pi /2$ known in electrodynamics as a wake effect 
(see, e.g., \cite{ryaz}). At the same time, it is strongly influenced by the special
geometry of mid-central collisions in such a way that the wake radiation at 
about $\pi /4$-orientation of the trigger particle can escape the overlap region
much easier than wake gluons for in- and out-of-plane orientations of the 
trigger particle.

Let me briefly remind that in macroscopic electrodynamics there are two
prominent collective effects induced by the polarization of the medium
due to a charge moving in it. Those are the transverse Cherenkov shock waves and
the longitudinal oscillations in the wake region behind the charge\footnote{
Unfortunately, both of them are often called as wakes.} (in distinction to
the transverse oscillations left behind the ship in hydrodynamics).

The wake effect follows from the equation
\begin{equation}
{\rm div} {\bf E}({\bf r}, \omega )=\rho ({\bf r}, \omega )/\epsilon (\omega ), 
\label{div1}
\end{equation}
where $\epsilon $ is the dielectric permittivity in electrodynamics.
Transforming this equation to the space-time coordinates
\begin{equation}
{\rm div} {\bf E}({\bf r}, t )=\rho ({\bf r}, t ) +\int _0^{\infty }d\tau 
\rho ({\bf r}, t-\tau )\int _{-\infty}^{\infty }\frac {d\omega }{2\pi }
\frac {1-\epsilon (\omega )}{\epsilon (\omega )}\exp (i\omega \tau )
\label{div2}
\end{equation}
one gets the additional contribution to the current density due to 
$1/\epsilon (\omega )$-term. It is clearly seen from (\ref{div2}) that 
this effect is determined by the collective properties of the medium because it 
vanishes in vacuum at $\epsilon =1$. Its contribution is especially important 
at those frequencies where $\epsilon (\omega )=0$, and in that case the medium
oscillations are strictly positioned along the way of the parton ($z$-axis)
\cite{ryaz}:
\begin{equation}
\rho ({\bf r}, t )=g\delta (x)\delta (y)\theta (t-z)\omega 
\sin [\omega (z-t)]\exp [-\delta (t-z)].
\label{wake}
\end{equation}
In principle, the dispersion of $\epsilon (\omega )$ is
important. These oscillations, however, exist also for $\epsilon (\omega )$ 
different from 0. Our goal here
is to find their strength and the angular distribution of the radiation
emitted due to these oscillations.

At the classical level the quark-gluon medium is described \cite{epj} by the
equations similar to those of electrodynamics with color indices added and
the chromopermittivity replacing the dielectric permittivity. For a 
relativistic parton moving in it along $z$-axis with a constant velocity $v$
the corresponding scalar ($\Phi $) and vector ($\bf A$) potentials in the
momentum space can be cast in the form 
\begin{equation}
\Phi_a^{(1)}=2\pi gQ_a\frac {\delta (\omega -kv\zeta)v^2\zeta ^2}{\omega ^2
\epsilon(\epsilon v^2\zeta ^2-1)},
\label{phi}
\end{equation}

\begin{equation}
A_{z,a}^{(1)}=\epsilon v\Phi_a^{(1)},
\label{a}
\end{equation}
\begin{equation}
\zeta=\cos \theta ,
\end{equation}
$g$ is the coupling constant, $\omega , k $ are the energy and momentum,
$\epsilon $ is the chromopermittivity\footnote{We use the same letter because
it would not lead to any confusion.}, $\theta $ is the polar angle. The
superscript (1) indicates the classical lowest order contribution. Let us
note that both the Cherenkov term $(\epsilon v^2\zeta ^2-1)^{-1}$  and
the wake factor $\epsilon ^{-1}$ are present.

The form of the spatio-temporal extent
of these potentials in a static and infinitely extended quark-gluon plasma was
considered in \cite{rm, cmt, cmrt} both for collisioneless case and with
collisions taken into account. Even though it provides intuitive insight for
the medium response to the parton motion, we are mostly interested  in its 
energy loss and the angular distribution of the corresponding radiation
observed in experiment.

The energy loss $dW$ per the length $dz$ is determined by the formula
\begin{equation}
\frac {dW}{dz}=-gE_z.    \label{eloss}
\end{equation}
In the lowest order
\begin{equation}
E_z^{(1)}=i\int \frac {d^4k}{(2\pi )^4}[\omega A_z^{(1)}({\bf k},\omega)-
k_z\Phi^{(1)}({\bf k},\omega)]e^{i({\bf k}{\bf v}-\omega)t}.
\label{ez}
\end{equation}
Inserting (\ref{phi}), (\ref{a}) in (\ref{ez}), (\ref{eloss}) one gets
\begin{equation}
\frac {dW^{(1)}}{dzd\zeta d\omega }=\frac {g^2\omega }{2\pi ^2v^2\zeta }
{\rm Im}\left(\frac {v^2(1-\zeta ^2)}{1-\epsilon _t v^2\zeta ^2}-
\frac {1}{\epsilon _l }\right),
\label{dwdz}
\end{equation}
The first term in the brackets corresponds to the transverse gluon Cherenkov
radiation (index $t$ at $\epsilon _t$) and the second term to the radiation
due to the longitudinal wake (index $l$ at $\epsilon _l$). The transverse and
longitudinal components of the chromopermittivity  tensor are explicitly
indicated here even though they are equal in any homogeneous medium.

Similar to electrodynamics \cite{gr}, one easily gets from (\ref{dwdz}) the 
energy-angular spectrum of emitted gluons \cite{dklv}
\begin{equation}
\frac {dN^{(1)}}{dzdx d\omega }=\frac {dW^{(1)}}{\omega dzd\zeta ^2 d\omega }=
\frac {\alpha _SC}{2\pi }\left [
\frac {(1-x)\Gamma _t}{(x-x_0)^2+(\Gamma _t)^2/4}+\frac {\Gamma _l}{x}\right ], 
\label{9}
\end{equation}
where 
\begin{equation}
x=\zeta ^2, \;\;\;  
x_0=\epsilon_{1t}/\vert \epsilon _t \vert ^2v^2, \;\;\;                         
\Gamma _j=2\epsilon_{2j}/\vert \epsilon _j \vert ^2v^2, \;\;\;
\epsilon _j=\epsilon _{1j}+i\epsilon _{2j}.
\end{equation}

The first term describes the gluon Cherenkov radiation with the imaginary 
part of $\epsilon $ taken into account. The conical emission pattern at the 
constant angle $\theta _0=\arccos \sqrt {x_0}$ typical for constant real $\epsilon $
is replaced by a'la Breit-Wigner angular shape. Namely this term was used
in \cite{dklv} to fit the experimental data about the two-hump structure in
central Au-Au collisions at 200 GeV. The information about  the real and 
imaginary parts of $\epsilon $ was obtained. Their values were somewhat 
different for STAR and PHENIX data because the hump positions differed in
the earlier data (for more details see \cite{dklv}). Since both collaborations 
agree now \cite{holz} that the peaks are positioned at 
$\Delta \phi \approx \pi \pm 1.1$ rad we use the approximate values of 
$\epsilon _{1t}
\approx 6$ and  $\epsilon _{2t} \approx 0.8$ in what follows. They are
slightly shifted between the earlier values for fits \cite{dklv} of STAR
and PHENIX data.

The second term in (\ref{9}) corresponds to the radiation induced by the
longitudinal oscillations in the wake left behind the away-side parton. It is
emitted mostly in the direction perpendicular to the wake. That reminds the
dipole radiation. Namely the color charge-changing regions of the dipole-like
structure behind the parton are seen in the Monte Carlo simulations of the 
wake \cite{cmt, cmrt}. The $1/x$-singularity is however much stronger than
the $\sin ^2\theta $ dipole distributions. It must be somehow saturated in
the vicinity of $x=0$. Therefore we use this term for qualitative conclusions 
in the region not very close to $x=0$. Replacing the angle $\theta $ counted
from the direction of the away-side jet by the angles in the laboratory
system $\theta _L,\; \phi _L \;\; (\cos ^2\theta =\sin ^2\theta _L\cos ^2\phi _L)$
one notices that $\cos ^2\phi _L \geq x =\cos ^2\theta $. To get the 
$\phi _L$-distribution one must integrate over $\theta _L$ as it was done
for Cherenkov radiation in \cite{dklv}.

To simplify estimates we, first, compare the wake radiation with Cherenkov
radiation at the peak of the latter $x\approx x_0$. Their ratio is 
($v^2\approx 1$)
\begin{equation}
\frac {\Gamma _t\Gamma _l}{4x_0(1-x_0)}\approx \frac {\epsilon _{2t}
\epsilon _{2l}}
{\epsilon _{1t}(1-\epsilon _{1t}/\vert \epsilon \vert ^2)} \approx
4\cdot 10^{-3}\ll 1.
\end{equation}
The numerical estimate is done for $\epsilon _t=\epsilon _l$ valid for the 
homogeneous medium. From it one concludes that the wake radiation is 
negligible at the peak of the Cherenkov humps. However, it strongly
increases at small $\pi-\Delta \phi _L$. They become comparable at
\begin{equation}
x_e\approx \frac {x_0^2}{2x_0+1}
\end{equation}
that corresponds to the angles $\pi-\Delta \phi _L\approx 1.43 $ rad.
The wake radiation overwhelms the Cherenkov contribution at 
$\Delta \phi _L<\pi -1.43$ rad where the latter decreases. Therefore the shift 
of their combined maximum to $\pi-\Delta \phi _L\approx 1.3$ rad noticed 
in \cite{holz} seems quite plausible.

To compute the absolute value of the intensity of the radiation one should 
take into account the geometry of the collisions. First of all, as follows from
Eq. (\ref{9}) it is proportional to the length of the trajectory of the
away-side parton inside the region where the colliding nuclei overlap. Second,
one must consider the lengths of the trajectories of the created gluons in
the medium. They are not taken into account in Eq. (\ref{9}) where the 
intensity of the radiation at the creation points is shown. It should be damped
exponentially with its path from the creation point to the surface. The 
interplay of these two factors determines the optimum trajectory for the
away-side parton radiation to escape the interaction region and be detected.

All these factors are inessential for central collisions. They would produce 
the overall normalization factor which is anyway left undetermined. We can only 
rely on the chance for the radiation to leave the overlap region with
detectable intensity. Once it has been already detected in experiment we are 
sure that this factor is not negligibly small. In central collisions the 
overlap region is spherically symmetric. If it is considered being at rest 
on the average with no strong internal flows then 
there is no difference at which angle is the trigger particle oriented. However,
the collision axis imposes some special direction in the problem which could
influence the chromopermittivity tensor Lorenz-transforming it. Therefore, 
even for central collisions it would be instructive to measure the Cherenkov 
humps for the trigger orientation different from about $\pi /2$ (see \cite{d}).

In mid-central collisions this region reminds the ellipsoidal rugby ball.
For 20-25\% events detected in \cite{holz} the ratio of its principal axes
can be as small as 3/4. The intensity of the detectable radiation strongly
depends on how long is the trajectory of the away-side parton inside the 
overlap region and how close it is to the surface. One should also take
into account that the trigger particle is created near the surface. Then using 
the rugby ball model one can show that for partons with trajectories oriented 
at the angles about $\pi /4$ just between the main (in- and out-) planes
the conditions for their radiation emitted perpendicular to their trajectories
is optimal compared to in- and out-of-plane orientations. Therefore the
effect observed by PHENIX collaboration in mid-central nuclear collisions
\cite {holz} is qualitatively explained as a result of the wake radiation.
To get quantitative comparison with the data one should compute it with the 
geometry of collisions and damping factors fully taken into account.

In conclusion, it is argued that the effect of the shifted maxima positions and
different intensity for
in(out)-of-plane radiation compared with that in between these planes detected
in the PHENIX experiment \cite{holz} is qualitatively explained as being
due to the wake radiation emitted at the trace of the away-side parton with 
the emission angles close to $\pi /2$ to the parton orientation.

\end{document}